\documentclass[conference,10pt,letterpaper]{IEEEtran}
\IEEEoverridecommandlockouts

\usepackage{cite}

\usepackage{array}
\usepackage{footnote}
\makesavenoteenv{tabular}
\usepackage{booktabs}

\usepackage[utf8]{inputenc}
\usepackage{graphicx}

\usepackage{amsmath,amssymb,amsfonts}
\usepackage{algorithmic}
\usepackage{textcomp}
\usepackage{xcolor}

\usepackage{hyperref}
\hypersetup{
    colorlinks = false, 
    linkcolor = black,  
    urlcolor = black,   
    citecolor = black   
}

\def\BibTeX{{\rm B\kern-.05em{\sc i\kern-.025em b}\kern-.08em
    T\kern-.1667em\lower.7ex\hbox{E}\kern-.125emX}}

\DeclareRobustCommand*{\IEEEauthorrefmark}[1]{%
    \raisebox{0pt}[0pt][0pt]{\textsuperscript{\footnotesize\ensuremath{#1}}}
}

\begin{document}

\title{OpenTCM: A GraphRAG-Empowered LLM-based System for Traditional Chinese Medicine Knowledge Retrieval and Diagnosis\\
\thanks{*Hongkai Chen is the corresponding author.}
}

\author{
	\IEEEauthorblockN{
		Jinglin He\IEEEauthorrefmark{1}, 
		Yunqi Guo\IEEEauthorrefmark{1}, 
		Lai Kwan Lam\IEEEauthorrefmark{2}, 
		Waikei Leung\IEEEauthorrefmark{3},
            Lixing He\IEEEauthorrefmark{1},} 
         \IEEEauthorblockN{   
            Yuanan Jiang\IEEEauthorrefmark{3},
            Chi Chiu Wang\IEEEauthorrefmark{2},
            Guoliang Xing\IEEEauthorrefmark{1},
            and Hongkai Chen*\IEEEauthorrefmark{1}
		 } 
	\IEEEauthorblockA{\IEEEauthorrefmark{1}Department of Information Engineering\\ The Chinese University of Hong Kong, HKSAR, China\\ Email: \{jinglinhe01, guoyunqi, helixing99\}@gmail.com, \{glxing, hkchen\}@ie.cuhk.edu.hk}
	\IEEEauthorblockA{\IEEEauthorrefmark{2}Department of Obstetrics and Gynaecology\\ The Chinese University of Hong Kong, HKSAR, China\\ Email: \{laikwanlam, ccwang\}@cuhk.edu.hk}
	\IEEEauthorblockA{\IEEEauthorrefmark{3}School of Chinese Medicine\\ The Chinese University of Hong Kong, HKSAR, China\\ Email: theresaleungwk@link.cuhk.edu.hk, yajiang@cuhk.edu.hk} 
}

\maketitle

\begin{abstract}
Traditional Chinese Medicine (TCM) represents a rich repository of ancient medical knowledge that continues to play an important role in modern healthcare. 
Due to the complexity and breadth of the TCM literature, the integration of AI technologies is critical for its modernization and broader accessibility. 
However, this integration poses considerable challenges, including the interpretation of obscure classical Chinese texts and the modeling of intricate semantic relationships among TCM concepts. 
In this paper, we develop OpenTCM, an LLM-based system that combines a domain-specific TCM knowledge graph and Graph-based Retrieval-Augmented Generation (GraphRAG). 
First, we extract more than 3.73 million classical Chinese characters from 68 gynecological books in the Chinese Medical Classics Database, with the help of TCM and gynecology experts. 
Second, we construct a comprehensive multi-relational knowledge graph comprising more than 48,000 entities and 152,000 interrelationships, using customized prompts and Chinese-oriented LLMs such as DeepSeek and Kimi to ensure high-fidelity semantic understanding. 
Last, we empower OpenTCM with GraphRAG, enabling high-fidelity ingredient knowledge retrieval and diagnostic question-answering without model fine-tuning.
Experimental evaluations demonstrate that OpenTCM achieves mean expert scores (MES) of 4.378 in ingredient knowledge retrieval and 4.045 in diagnostic question-answering tasks, outperforming state-of-the-art solutions in real-world TCM use cases.
\end{abstract}


\section{Introduction}

Traditional Chinese Medicine (TCM) represents a rich repository of ancient medical knowledge that continues to hold significant value in modern healthcare systems~\cite{TCM}. 
Due to the complexity of the literature and the practice of TCM, many researchers have tried to build systems to assist with consultation and diagnosis for TCM physicians~\cite{Biancang,lingdan}. 
Current TCM knowledge retrieval and diagnosis systems predominantly often follow a paradigm of fine-tuning general-purpose Large Language Models (LLMs), such as Llama~\cite{touvron2023llama}, on domain-specific datasets. 
However, they often face significant challenges such as the complexity of parsing classical Chinese texts, the intricate relationships among TCM concepts (such as ingredients, symptoms, and diagnostic principles), and the absence of structured knowledge such as knowledge graphs~\cite{TCMAI1,TCMAI3}.
In particular, ancient TCM literature is often written in archaic language, making them difficult for modern practitioners and computational models to interpret. Furthermore, existing approaches primarily focus on modern medical literature, overlooking the foundational corpus of classical TCM~\cite{medicalGraphRAG}. 
Although recent advances in LLMs have proven to be effective in medical knowledge democratization, the direct application of general-purpose LLMs to TCM still faces significant limitations, such as hallucination in prescription generation and semantic understanding of classical Chinese.

To address these limitations, we propose \textit{OpenTCM}, an LLM-based TCM knowledge retrieval and diagnosis system that integrates domain-specific  knowledge with Retrieval-Augmented Generation with Graphs~(GraphRAG)~\cite{GraphRAG4}. 
First, we extract over 3.73 million ancient Chinese characters from 68 gynecological books sourced from the Chinese Medical Classics Database~\cite{TCMcollections}, with verification from experienced TCM and gynecology experts. 
Next, we construct a comprehensive multi-relational knowledge graph containing over 48,000 entities and 152,000 interrelationships. This graph spans key TCM concepts, including 3,700+ ingredients, 14,000+ diseases, 17,000+ symptoms and treatments, and 65,000+ ingredient references. 
The construction process leverages customized prompts and domain-adapted LLMs, such as DeepSeek~\cite{liu2024deepseek} and Kimi~\cite{team2025kimi}, to ensure accurate semantic understanding of classical Chinese medical literature. 
Compared to general-purpose prompts and LLMs, our approach achieves significant improvements in knowledge graph quality, with a precision of 98.55\%, recall of 99.60\%, F1-score of 99.55\%, and accuracy of 98.17\%.
In addition, our experiments show that OpenTCM achieves a Mean Expert Score (MES), accuracy, and inter-rater agreement (IRA) of 4.378, 99.0\%, and 0.057 for ingredient retrieval, and 4.045, 98.8\%, -0.013 for diagnostic question-answering, respectively.
By eliminating the need for model fine-tuning, OpenTCM leverages retrieval-augmented generation to provide accurate, context-aware responses with minimal computational overhead. 
Our structured knowledge graph also mitigates hallucination risks and enhances reasoning over complex TCM relationships, offering a scalable and efficient alternative to traditional methods.

Our main contributions are summarized as follows.
\begin{itemize}
\item We develop \textit{OpenTCM}, the first LLM-based TCM knowledge retrieval and diagnosis system that integrates a large corpus of structured TCM knowledge derived from ancient TCM literature, which consists of 68 gynecological books with over 3.73 million ancient Chinese characters from the Chinese Medical Classics Database.
\item We propose a TCM knowledge graph construction approach that combines customized prompts with domain-specific LLMs, achieving high-fidelity semantic extraction from classical texts.
\item We integrate the knowledge graph with GraphRAG to enhance reasoning capabilities over complex TCM interrelationships while maintaining computational efficiency.
\item Our experimental evaluation demonstrates that our system significantly outperforms existing solutions, achieving a precision of 98.55\% in knowledge graph construction task, an MES of 4.378 and accuracy of 99.0\% in ingredient retrieval, and an MES of 4.045 and accuracy of 98.8\% in diagnostic question-answering.
\end{itemize}

The remainder of this paper is organized as follows: \autoref{sec:related_work} reviews related work on TCM knowledge graphs and LLMs. \autoref{sec:system_architecture} introduces our system architecture, including data collection, knowledge graph construction, and the GraphRAG framework. \autoref{sec:experiments} presents experimental evaluations. Finally, \autoref{sec:future_work} outlines future work and \autoref{sec:conclusion} concludes our work.
Our project will be publicly available at \url{https://github.com/XY1123-TCM/OpenTCM}.

\section{Related Work}
\label{sec:related_work}
This section reviews the existing literature on TCM knowledge graphs, training-based LLMs for TCM, and the emerging fusion of knowledge graphs with LLMs via GraphRAG. 
\subsection{TCM Knowledge Graphs}
Knowledge graphs have emerged as a promising tool to structure TCM’s complex knowledge system. For example, Zhang et al.~\cite{electronics13071395} utilize LLMs for TCM knowledge graph construction, employing named entity recognition and few-shot learning to reduce manual annotation efforts in applications such as education, diagnosis, and treatment. 
Similarly, Duan et al. \cite{article} propose a hybrid method integrating LLMs with manual verification to build a TCM case knowledge graph, enabling a question-answering system. 
While these efforts demonstrate the potential of knowledge graphs in TCM, they do not incorporate the capacity to reason over relational complexities---a gap OpenTCM aims to address. 
In contrast, we incorporate GraphRAG in our OpenTCM system. enabling reasoning over relational complexities in TCM literature.

\subsection{LLMs for TCM}
In recent years, driven by the success of LLMs in modern medical domains~\cite{fu2025shadead}, the application of LLMs for TCM has attracted significant attention to enhance clinical reasoning and knowledge access. 
BianCang~\cite{Biancang}, built on the Qwen-2/2.5 architecture, undergoes continuous pre-training and fine-tuning to embed TCM expertise, improving its domain-specific performance. 
Lingdan~\cite{lingdan}, based on Baichuan2-13B-Base, employs a Chain-of-Thought approach, achieving an 18.39\% improvement in Top20 F1-score for symptom analysis and prescription tasks compared to baseline solutions. 
Other models, such as Qibo~\cite{qibo} and Zhongjing~\cite{zhongjing}, further showcase LLM's potential in TCM consultations, with matching capabilities to ChatGPT with fewer parameters. 
TCMChat~\cite{tcmchat} uses pre-training and supervised fine-tuning on curated TCM text and Chinese question-answer datasets, focusing on tasks like entity extraction and ingredient recommendation. 
These studies, however, primarily rely on resource-intensive fine-tuning and lack structured knowledge integration, often resulting in computational inefficiency and hallucinations.
In contrast, OpenTCM uses GraphRAG to realize high fidelity without computationally expensive pre-training or fine-tuning. 

\subsection{Knowledge Graph-based Medical LLM}
Recently, fusion of knowledge graphs with LLMs via GraphRAG has explored in medicine. 
The most related paper to our work is MedGraphRAG~\cite{medicalGraphRAG}, which proposes a graph-based RAG framework for general medicine, outperforming state-of-the-art models across multiple benchmarks. 
However, its focus on modern medicine makes TCM-specific knowledge retrieval and diagnosis challenging. 
In this work, OpenTCM employs a training-free approach using GraphRAG, which not only reduces computational overhead but also integrates the rich relational data from the TCM literature to improve response accuracy and contextual relevance.

\begin{figure*}[htp]
    \centering
    \includegraphics[width=\textwidth]{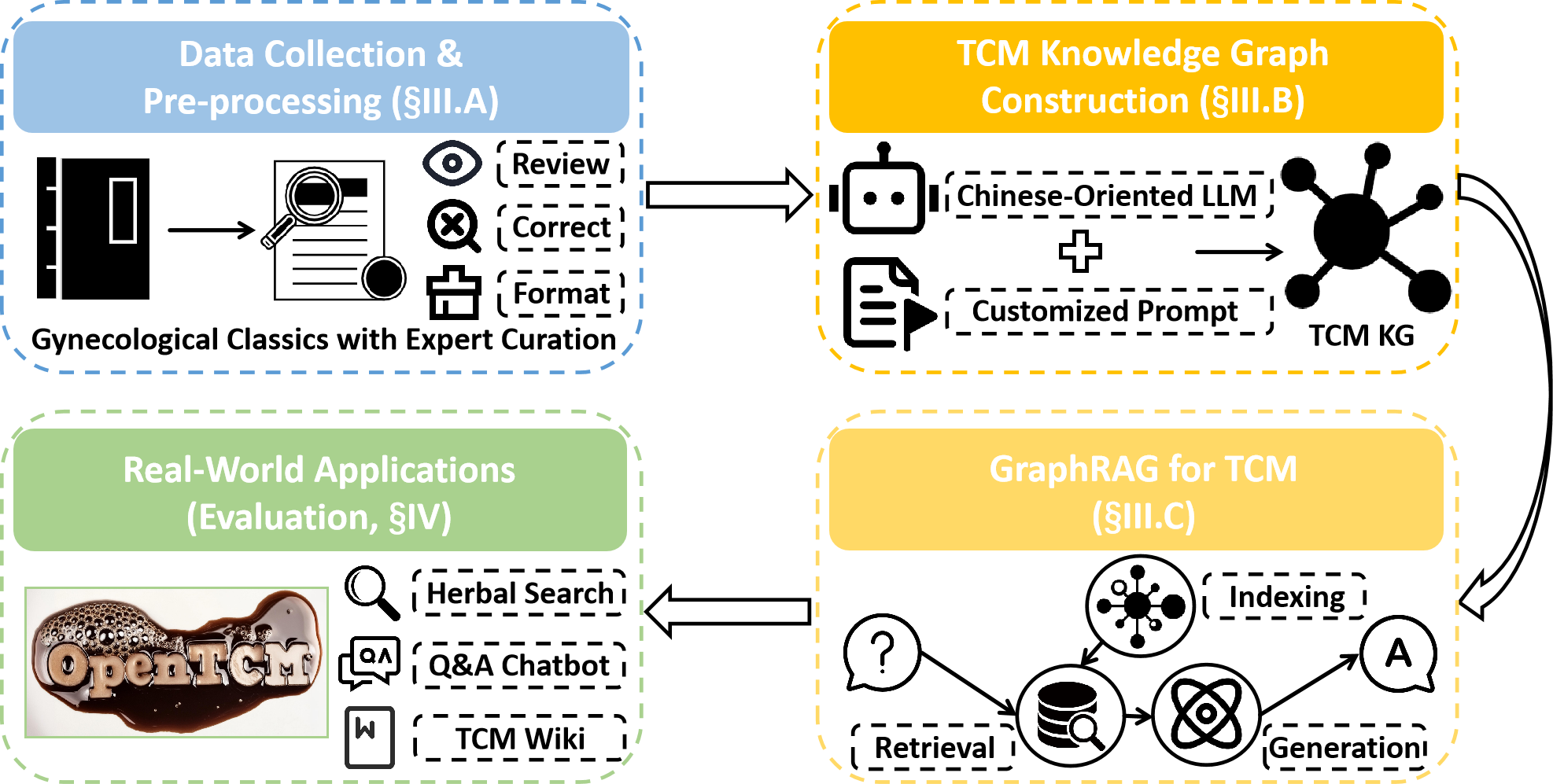}
    \caption{System architecture of OpenTCM, which consists of three main components: (1)~data collection and pre-processing, (2)~knowledge graph construction, and (3)~graph-based retrieval-augmented generation.}
    \label{fig:system}
\end{figure*}

\section{System Overview}
\label{sec:system_architecture}
In this section, we will describe the system overview of the OpenTCM.
As shown in Figure~\ref{fig:system}, OpenTCM consists of three main components: (1)~data collection and pre-processing, (2)~knowledge graph construction, and (3)~graph-based retrieval-augmented generation. Real-world applications are detailed in \autoref{sec:experiments}.

The process starts with collecting gynecological books from the Chinese Medical Classics database~\cite{TCMcollections}. 
These medical literature, covering obstetrics, gynecology, and fertility, are carefully reviewed and corrected by TCM and gynecology medical experts to ensure data quality.
Next, a multi-relational knowledge graph is constructed using well-customized prompts and advanced domain-adapted LLMs like Deepseek and Kimi. 
This knowledge graph captures key TCM elements such as ingredients, symptoms, and treatments, forming a structured representation of TCM.
Last, the constructed knowledge graph serves as the backbone for the database, enabling the implementation of GraphRAG. 
This integration allows OpenTCM to offer powerful features such as TCM ingredient search and diagnostic question-answering without requiring pre-training or fine-tuning of the LLMs, significantly reducing computational overhead while maintaining high performance.

\subsection{Data Collection and Pre-processing}
We extract 68 ancient classical books on traditional Chinese medicine in gynecology (20 Obstetrics, 43 Gynecology, and 5 Fertility) from the Digital Resource Library of Ancient Chinese Medicine Ancient Books of China, the most authoritative and systematic database of research on traditional Chinese medicine \cite{TCMcollections}.
The books contain 6,787 chapters and 3,731,358 characters, covering diverse topics such as prescriptions, traditional Chinese patent medicines, simple preparations and medical case references (see Tables \ref{Statistics on the Quantity of TCM book and character} and \ref{Statistics on the Quantity of information in TCM Database}).
After collection, these books are then handed over to experienced TCM practitioners from Traditional Chinese Medicine and Gynecology for rigorous review, correction, and formatting to produce clean, usable dataset.
 
\begin{table}[!htbp]
\caption{Number of books, chapters, and characters in different TCM specialties}
\label{Statistics on the Quantity of TCM book and character}
\begin{center}
\begin{tabular}{ >{\centering\arraybackslash}m{1.6cm} 
>{\centering\arraybackslash}m{1.2cm} 
>{\centering\arraybackslash}m{1.2cm} 
>{\centering\arraybackslash}m{1.2cm}
>{\centering\arraybackslash}m{1.2cm}}
\toprule  
Quantity & Obstetrics  &  Gynecology &  Fertility &  Total\\ 
\midrule  
Book & 20 & 43 & 5 & 68\\
Chapter & 1987 & 4496 & 304 & 6787 \\
Character & 734095 & 2813900 & 183363 & 3731358\\
\bottomrule  
\end{tabular}
\end{center}
\end{table}

\begin{table}[!htbp]
\caption{Number of ingredients, diseases, symptoms, treatments and ingredient references in TCM Database}
\label{Statistics on the Quantity of information in TCM Database}
\begin{center}
\begin{tabular}{ >{\centering\arraybackslash}m{1.3cm} 
>{\centering\arraybackslash}m{1.3cm} 
>{\centering\arraybackslash}m{1.3cm}
>{\centering\arraybackslash}m{1.3cm}
>{\centering\arraybackslash}m{1.3cm}
}
\toprule  
Ingredients& Diseases & Symptoms  & Treatments & Ingredient-References\\ 
\midrule  
3737 & 14059 & 17031 & 17031 & 65847\\
\bottomrule  
\end{tabular}
\end{center}
\end{table}

\subsection{Knowledge Graph Construction}
We use LLMs to extract structured knowledge graphs from a large corpus of classical Chinese medical texts. 
This process faces two primary challenges: (1)~accurately interpreting ancient Chinese medical language and (2)~extracting high-quality, structured information such as diseases, symptoms, treatments, and herb references. 
To address these challenges, we leverage Chinese-centric LLMs and carefully designed prompts. 
Our final knowledge graph includes 48{,}406 entities and 152{,}754 relations, among which 10 types of triplets are formed, as depicted in Table~\ref{Constructed Interrelationships}..

The core of our approach lies in the design of an effective system prompt. Below, we summarize its key components:

\begin{itemize}
    \item \textbf{Role Specification}: The prompt defines the model as a ``TCM knowledge analysis assistant'', setting expectations for professional, structured, and accurate output.

    \item \textbf{Context Injection}: Metadata such as book name, chapter name, and chunk index are included to enhance reference resolution and enable extraction of meta-relations like ``Belong to Book''.

    \item \textbf{Task Instructions}: The model is explicitly instructed to extract only 10 predefined relation types (e.g., ``Treat Disease,'' ``Use Ingredient'') and return results as a list of (subject, predicate, object) triples.

    \item \textbf{Output Format Constraints}: The output must strictly conform to a JSON array of objects, each containing subject, predicate, and object. Invalid or ambiguous content is to be skipped.

    \item \textbf{Few-shot Demonstration}: An example passage from a TCM classic and its corresponding JSON-formatted triples are provided to illustrate expected behavior and reinforce output format.
\end{itemize}
\begin{table}[t]
\caption{Constructed Interrelationships of TCM Knowledge Graph}
\label{Constructed Interrelationships}
\begin{center}
\begin{tabular}{ >{\centering\arraybackslash}m{2.5cm} 
>{\centering\arraybackslash}m{0.8cm} 
>{\centering\arraybackslash}m{3.2cm} 
>{\centering\arraybackslash}m{0.8cm} 
}
\toprule  
Interrelationship & Num &Interrelationship&Num\\ 
\midrule  
Belong to Category & 48406 &Include Section&294\\ 
Include Chapter & 6786 &Belong to book&6786\\ 
Treatment Plan& 17001 &Treat Disease&16133\\ 
Describe Disease & 16104 &Treatment Symptom&13605\\ 
Symptoms Present & 13581 &Ingredient Use&65846\\ 
\bottomrule  
\end{tabular}
\end{center}
\end{table}

\begin{figure}[htp]
    \centering
    \includegraphics[width=\linewidth]{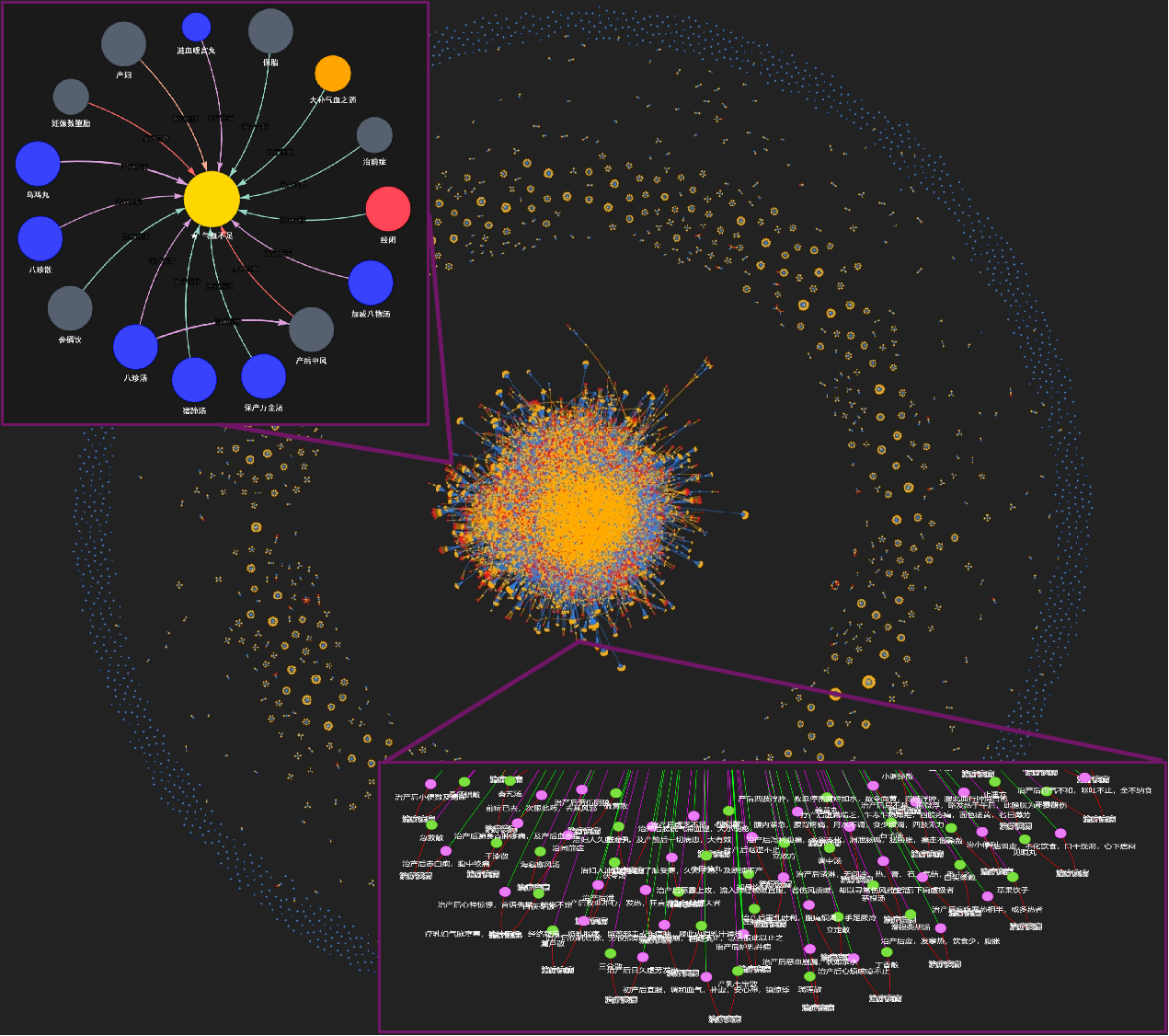}
    \caption{Visualization of our TCM knowledge graph.}
    \label{fig:TCM KG}
\end{figure}

This prompt design effectively guides LLMs to extract structured data from unstructured ancient text, ensuring high consistency, interpretability, and ease of downstream processing for large-scale knowledge graph construction.
The whole TCM Knowledge Graph and two partial subgraphs are visualized in Figures~\ref{fig:TCM KG}.
This structured knowledge graph not only preserves the richness of TCM data but also facilitates advanced analysis and application in modern research contexts.

\begin{table}[t]
\caption{Performance of customized prompt verses general prompt, KIMI-based}
\label{Performance of customized prompt verses general prompt, KIMI-based}
\begin{center}
\begin{tabular}{ >{\centering\arraybackslash}m{2.5cm} 
>{\centering\arraybackslash}m{2.5cm} 
>{\centering\arraybackslash}m{2.5cm} 
}
\toprule  
 & General & Customized\\ 
\midrule  
Precision & 90.1\% & \textbf{98.55\%}\\
Recall & 95.9\% & \textbf{99.60\%}\\
F1-score & 92.3\% & \textbf{99.55\%}\\
Accuracy & 86.8\% & \textbf{98.17\%}\\
\bottomrule  
\end{tabular}
\end{center}
\end{table}

\subsection{Graph-based Retrieval-Augmented Generation}

We use GraphRAG~\cite{graphRAG,graphRAG3,GraphRAG4} to integrate graph-structured data into OpenTCM to enhance retrieval and generation processes. 
While conventional RAG approaches predominantly processes unstructured text or images, GraphRAG leverages the rich structural relationships embedded within a knowledge graph to improve both information access and reasoning capabilities. 
This is particularly valuable for TCM, which features complex inter-dependencies among ingredients, symptoms, treatments, and syndromes.

We adopt GraphRAG to operate effectively on the TCM knowledge graph by directly utilizing the multi-relational structure that is characteristic of traditional Chinese medicine theory. 
The knowledge graph includes over 48,000 entities and 152,000+ relationships such as \textit{treats}, \textit{includes}, \textit{associated with}, and \textit{belongs to}, which encode rich semantic information. 
By grounding retrieval in this structured graph rather than text, our system is able to more precisely trace clinically meaningful paths. 
For example, linking a symptom to a treatment through an associated ingredient or syndrome. This structure-aware approach allows the system to avoid irrelevant content and retrieve only contextually valid information, which is essential in domains where precise distinctions matter.
Moreover, GraphRAG’s ability to traverse and aggregate information along multi-hop paths (e.g., symptom $\rightarrow$ syndrome $\rightarrow$ treatment $\rightarrow$ ingredient) enables OpenTCM to model the layered reasoning inherent in TCM practice. 
Instead of treating symptoms and ingredients as isolated keywords, GraphRAG interprets user queries within the relational graph, thereby capturing nuanced dependencies among concepts that are typically missed by flat retrieval systems.

These adaptations enhances our system's performance by relying on a domain-specific graph structure.
Therefore, OpenTCM mitigates hallucination and noise commonly observed in generic LLM outputs. This design supports high-precision tasks such as ingredient knowledge retrieval (accuracy: 99.0\%) and diagnostic question-answering (accuracy: 98.8\%), while reducing the need for model fine-tuning. 
The integration of GraphRAG with the TCM knowledge graph thus forms the core of OpenTCM’s ability to deliver accurate, interpretable, and context-aware responses in real-world medical scenarios.

\begin{figure}[t]
    \centering
    \includegraphics[width=8cm, keepaspectratio]{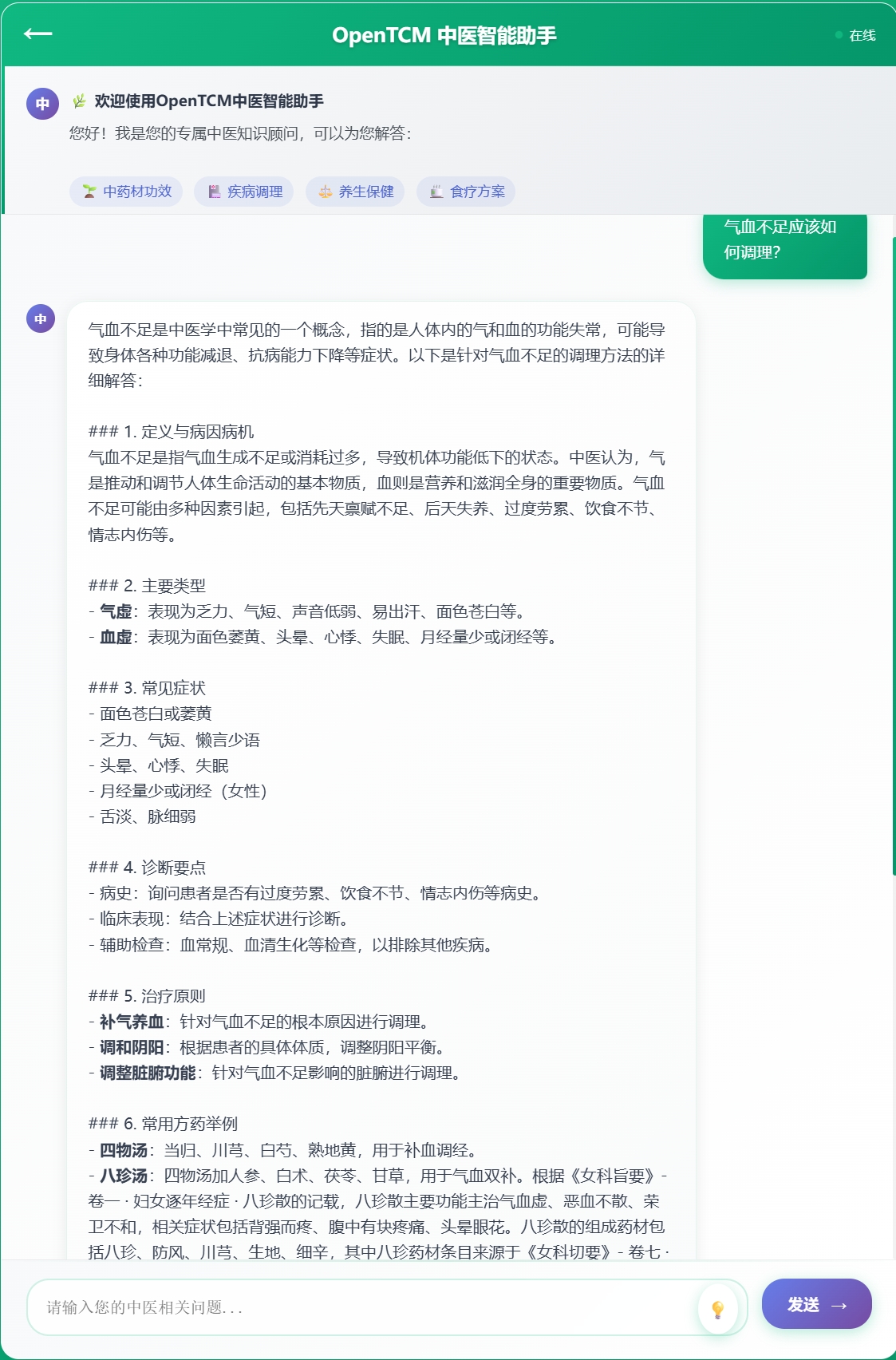}
    \caption{An example case of OpenTCM diagnostic question-answering.}
    \label{fig:QA}
\end{figure}

\section{Experimental Evaluation}
\label{sec:experiments}
To evaluate OpenTCM, we conduct a comprehensive experimental evaluation focusing on three key capabilities: knowledge graph construction, TCM ingredient knowledge retrieval, and diagnostic question-answering. 

\subsection{Baseline Models}
To ensure a thorough evaluation, we compare OpenTCM against several baseline models in both knowledge graph construction and real-world applications.

\subsubsection{Knowledge Graph Construction Baselines} \begin{itemize} \item {General-Purpose LLMs}: We utilize ChatGPT-4 \cite{chatgpt4} and Claude-2 \cite{wu2023comparative} as representatives of general-purpose large language models. \item {Domain-Specific LLMs}: Deepseek \cite{liu2024deepseek} and Kimi \cite{team2025kimi} are employed as domain-adapted Chinese LLMs, fine-tuned on medical and classical Chinese corpora. \end{itemize}

\subsubsection{Real-World Application Baselines} \begin{itemize} \item {General-Purpose LLMs}: ChatGPT-4 \cite{chatgpt4} and Claude-2 \cite{wu2023comparative} are used to assess the performance of general-purpose models in TCM applications. 
\item {TCM-Specific LLMs}: BianCang \cite{Biancang} is a recent LLM specifically designed for TCM tasks, incorporating domain-specific knowledge and training data. \end{itemize}

\subsection{Knowledge Graph Construction}
Evaluating the quality of the TCM knowledge graph is a cornerstone of this study, given its role as the foundation for OpenTCM’s downstream applications. 
In the absence of a pre-existing gold standard for TCM knowledge graph construction, we adopt a sampling inspection approach, enlisting 4 domain experts in TCM and Gynecology to assess the generated triples. 
Our evaluation examines two dimensions: (1) the effectiveness of customized versus general prompts, and (2) the performance of different backbone models, including Chinese-oriented LLMs (i.e., Deepseek~\cite{liu2024deepseek}, Kimi~\cite{team2025kimi}) and more general LLMs (i.e., ChatGPT-4~\cite{chatgpt4}, Claude~\cite{wu2023comparative}).

We task the LLMs with extracting 152,754 triples from a sampled subset of the corpus, comprising randomly selected chapters from the 68 gynecological books. In summary, the constructed TCM knowledge graph, containing over 48,000 entities and 152,000 relationships, is validated through expert review of randomly selected 600 chapters. To evaluate the knowledge graph of OpenTCM quantitatively, we randomly select 1795 triples with corresponding expert annotations.

\subsubsection{Metric}
This data enables a quantitative assessment using four standard metrics:
\begin{itemize}
\item {Precision}: The proportion of generated triples that are correct, reflecting the accuracy of the extracted knowledge.
\begin{equation}
    \text{Precision} = \frac{\text{TP}}{\text{TP} + \text{FP}}
\end{equation}
\item {Recall}: The proportion of actual correct triples successfully extracted by the LLM, indicating its coverage of the true knowledge.
\begin{equation}
    \text{Recall} = \frac{\text{TP}}{\text{TP} + \text{FN}}
\end{equation}
\item {F1-Score}: The harmonic mean of Precision and Recall, providing a balanced measure of extraction quality.
\begin{equation}
    \text{F1-Score} = 2 \cdot \frac{\text{Precision} \cdot \text{Recall}}{\text{Precision} + \text{Recall}}
\end{equation}
\item {Accuracy}: The proportion of correctly identified triples relative to the total ground truth, gauging overall correctness (though less commonly emphasized in generation tasks).
\begin{equation}
    \text{Accuracy} = \frac{\text{TP}}{\text{TP} + \text{FP} + \text{FN}}
\end{equation}
\end{itemize}
Here, TP (True Positives) denotes triples generated by the LLM and validated as correct by experts, FP (False Positives) represents incorrect triples generated by the LLM, and FN (False Negatives) indicates correct triples missed by the LLM.

\subsubsection{Performance}
We design specific prompts tailored to TCM data extraction, covering chapter information, treatment plans, ingredient details, and their relationships, which is more suitable to our task than the general-purpose prompt.
The customized prompts demonstrate a clear advantage over general prompts (as shown in Table \ref{Performance of customized prompt verses general prompt, KIMI-based}, KIMI-based), indicating their effectiveness in enhancing extraction accuracy and completeness.
Besides, our tailored prompt is compatible with multiple LLM from different vendors, including Deepseek, Kimi, GPT4 and Claude2. We observe that DeepSeek and Kimi show superior performance as shown in Table \ref{Performance of GPT-4, Claude2, Deepseek, KIMI}, we hypothesize that those models are more capable because they use more Chinese content as the training set.
This significant difference highlights the importance of using models specifically adapted to the linguistic and contextual characteristics of TCM literature.

\begin{table}[t]
\caption{Performance Comparison among Different backbone LLMs in Knowledge Graph Construction}
\label{Performance of GPT-4, Claude2, Deepseek, KIMI}
\begin{center}
\begin{tabular}{ >{\centering\arraybackslash}m{1.8cm} 
>{\centering\arraybackslash}m{1.2cm} 
>{\centering\arraybackslash}m{1.2cm} 
>{\centering\arraybackslash}m{1.2cm} 
>{\centering\arraybackslash}m{1.2cm}
}
\toprule  
 & GPT4 & Claude2 & Deepseek & KIMI\\ 
\midrule  
Precision& 94.6\% &94.26\%& \textbf{98.61\%} & 98.55\%\\
Recall & 98\% & 97.41\% & 99.27\% & \textbf{99.60\%}\\
F1-score & 95.6\% & 95.37\% & 98.49\% & \textbf{99.55\%}\\
Accuracy & 92.8\% & 91.96\% & 97.9\% & \textbf{98.17\%}\\

\bottomrule  
\end{tabular}
\end{center}
\end{table}

An expert review of the full knowledge graph, encompassing over 48,000 entities and 152,000 relationships, confirms its fidelity in capturing TCM’s intricate knowledge structure. The combination of customized prompts and domain-adapted LLMs not only ensures high-quality triple extraction—as evidenced by the Precision, Recall, F1-Score, and Accuracy but also establishes a solid foundation for OpenTCM’s downstream applications like TCM ingredient search and diagnostic question-answering functionalities.

\subsection{Real-world Applications}
To assess OpenTCM’s effectiveness on real-world applications, we evaluate its performance on two critical downstream tasks: ingredient medicine knowledge retrieval and diagnostic question answering. 
These tasks test the system’s ability to leverage the TCM knowledge graph and GraphRAG technology to provide accurate, contextually relevant, and user-friendly responses.

Due to the absence of a definitive ground truth, we collect and annotate a dataset consisting of 257 ingredient information search queries and 303 diagnostic consultation questions. 
Each query is processed by OpenTCM and several baseline systems, including general-purpose LLMs (KIMI) and recently proposed TCM-specific LLMs (BianCang). 
The system's responses were evaluated by a panel of four medical experts, comprising two clinicians with more than 20 years of clinical experience and two medical Ph.D. students whose assessments were validated by their respective supervisors. A 5-point Likert scale, ranging from 1 (irrelevant) to 5 (highly relevant), was employed for scoring.

\subsubsection{Metrics}
We employ three key metrics for evaluating the performance of the system:
\begin{itemize}
    \item \textbf{Mean Expert Score (MES)}: This metric represents the average score assigned by domain experts to assess the relevance and correctness of the retrieved or generated information. MES quantitatively evaluates the experts' subjective judgment of the system's response quality.
    \item \textbf{Accuracy}: Accuracy is defined as the proportion of expert-labeled correct responses out of the total number of queries. For the purpose of this evaluation, a response is deemed correct if its expert score is equal to or greater than 3. This metric measures the overall capability of the model to produce acceptable responses.
    \item \textbf{Inter-Rater Agreement (IRA)}: IRA quantifies the consistency among multiple domain expert evaluators to ensure the reliability of the assessment results. For the calculation of IRA, expert ratings are binarized: scores greater than or equal to 3 are categorized as "correct" (or "agreement"), while scores less than 3 are categorized as "incorrect" (or "disagreement"). IRA reflects the objectivity of the evaluation by measuring the extent to which evaluators agree on these binary classifications.
\end{itemize}

\subsubsection{Ingredient Knowledge Retrieval}
This task evaluates the model’s ability to return accurate and comprehensive information about ingredients such as “Danggui” and “Renshen”.

\begin{figure*}[htp]
    \centering
    \includegraphics[width=.8\textwidth, keepaspectratio]{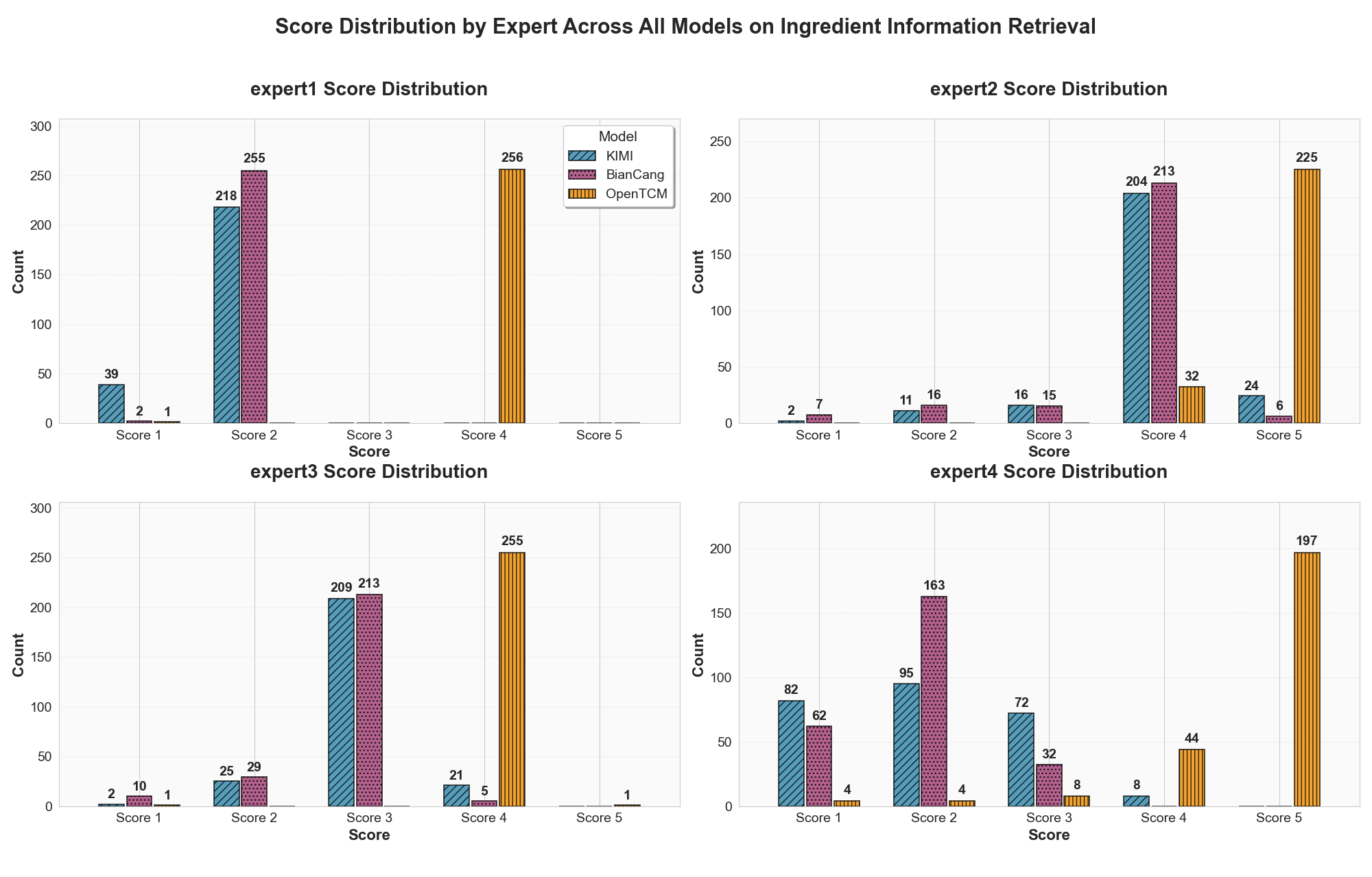}
    \caption{Score distribution of different models for Ingredient Knowledge Retrieval across four experts. OpenTCM shows significantly higher concentration at Score 4 and 5.}
    \label{fig:herb-dist}
\end{figure*}

Experimental results are presented in Table~\ref{tab:herb-retrieval} and Figure~\ref{fig:herb-dist}. OpenTCM achieves the highest Mean Expert Score (4.378), surpassing both general-purpose models and TCM-specific baselines. This demonstrates the effectiveness of knowledge graph-enhanced reasoning in improving retrieval precision and contextual relevance.

\begin{table}[t]
\caption{Performance Comparison among OpenTCM and baseline solutions in Ingredient Knowledge Retrieval}
\label{tab:herb-retrieval}
\begin{center}
\begin{tabular}{ 
>{\centering\arraybackslash}m{3cm} 
>{\centering\arraybackslash}m{1.3cm} 
>{\centering\arraybackslash}m{1.3cm} 
>{\centering\arraybackslash}m{1.3cm}
}
\toprule
Model & MES & Accuracy & IRA \\
\midrule
KIMI & 2.691  & 53.9\% & -0.154 \\
BianCang & 2.616 & 47.1\% & -0.166 \\
\textbf{OpenTCM} & \textbf{4.378} & \textbf{99.0\%} & \textbf{0.057} \\
\bottomrule
\end{tabular}
\end{center}
\end{table}

\subsubsection{Diagnostic Question Answering}
This task assesses the system’s diagnostic reasoning capabilities in clinical consultation scenarios (e.g., “How to treat menstrual disorders?” or “How to treat Qi deficiency with dampness?”). 
An example is illustrated in Figure~\ref{fig:QA}.

\begin{figure*}[htp]
    \centering
    \includegraphics[width=.8\textwidth, keepaspectratio]{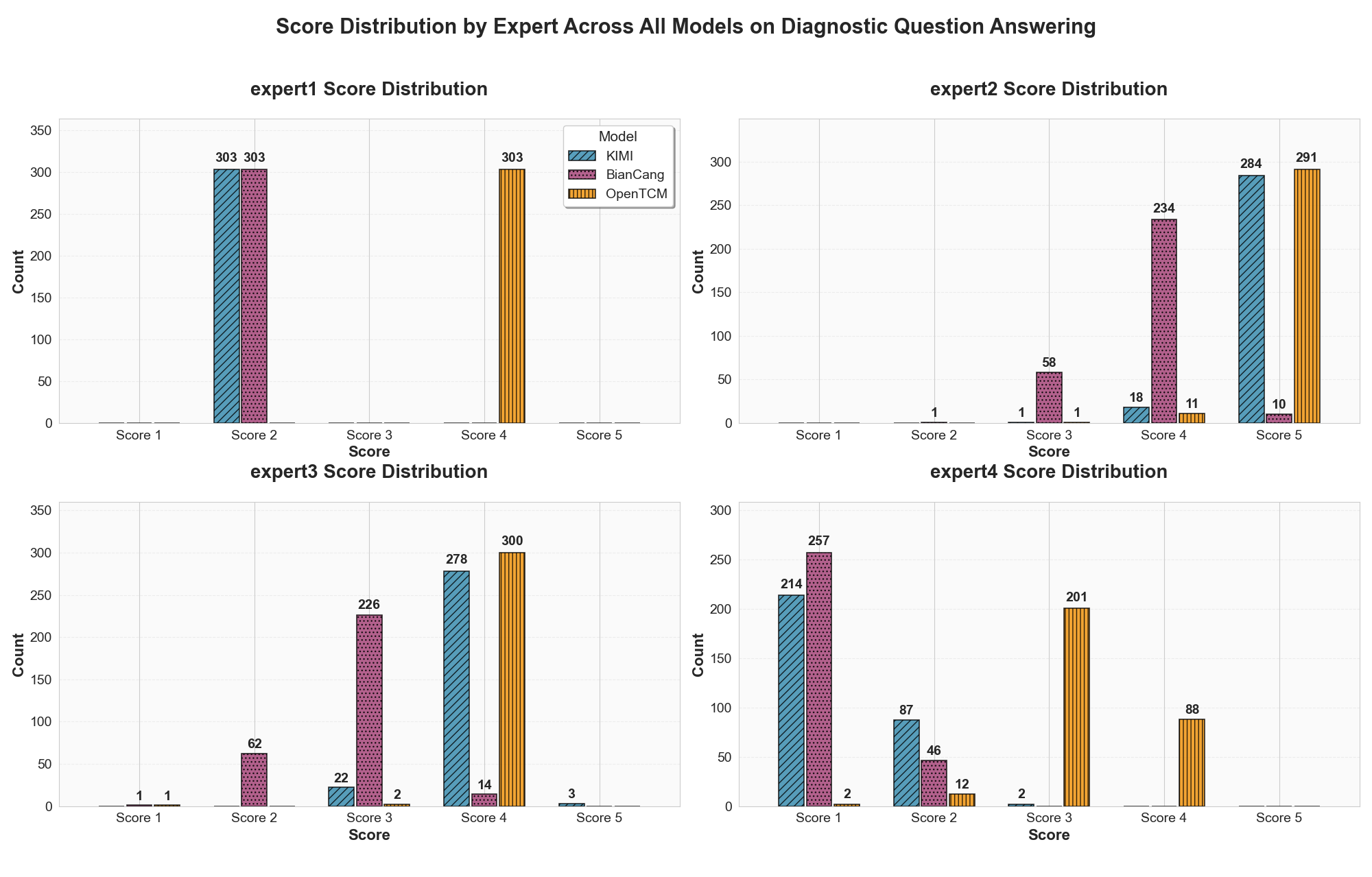}
    \caption{Score distribution of different models for Diagnostic Question Answering across four experts. OpenTCM responses receive notably higher expert ratings.}
    \label{fig:qa-dist}
\end{figure*}

Results in Table~\ref{tab:diagnostic-qa} and Figure~\ref{fig:qa-dist} show that OpenTCM maintains strong performance across all metrics, particularly in interpretability and contextual coherence, outperforming general LLMs and remaining competitive with TCM-specific models.

\begin{table}[!htbp]
\caption{Performance Comparison among OpenTCM and baseline solutions in Diagnostic Question-Answering}
\label{tab:diagnostic-qa}
\begin{center}
\begin{tabular}{ 
>{\centering\arraybackslash}m{3cm} 
>{\centering\arraybackslash}m{1.3cm} 
>{\centering\arraybackslash}m{1.3cm} 
>{\centering\arraybackslash}m{1.3cm}
}
\toprule
Model & MES & ACC & IRA \\
\midrule
KIMI & 3.043 & 50.2\%   &  -0.331  \\
BianCang & 2.455 & 44.7\%  &    -0.275 \\
\textbf{OpenTCM} & \textbf{4.045} & \textbf{98.8\%} & \textbf{-0.013} \\
\bottomrule
\end{tabular}
\end{center}
\end{table}


These results affirm OpenTCM’s practical value in real-world clinical and educational scenarios. Compared to both general-purpose and TCM-specific LLMs, OpenTCM exhibits superior performance in both factual accuracy and expert-perceived relevance, owing to its integration of structured domain knowledge through GraphRAG.

\section{Future Work}
\label{sec:future_work}
Future work will focus on expanding the TCM Knowledge Graph to include additional domains, such as rare diseases and historical case studies, to enhance OpenTCM's applicability. We also aim to refine the GraphRAG framework to support more complex queries and integrate multimodal data, such as image recognition for raw and dried botanic and non-botanic substances. Additionally, developing tailored interfaces for practitioners, researchers, and patients will improve accessibility to TCM knowledge. Finally, we plan to collaborate with medical practitioners and institutions to integrate OpenTCM into healthcare systems such as in-home elderly care~\cite{li2024emomarker}, clinical diagnostic~\cite{drhouse}, and exercise medicine~\cite{Retcare,Myotrainer_Demo}.

\section{Conclusion}
\label{sec:conclusion}
OpenTCM represents a significant advancement in applying LLMs to Traditional Chinese Medicine. By combining a meticulously constructed knowledge graph with GraphRAG, it achieves high accuracy and efficiency in knowledge retrieval and diagnostic tasks, paving the way for broader adoption of TCM practices.

\section*{Acknowledgment}
This work is partly supported by Chinese Medicine Development Fund, Hong Kong SAR, China under 23B2/034A\_R1.

\bibliographystyle{IEEEtran}  
\bibliography{reference.bib}{}

\begin{thebibliography}{10}
\providecommand{\url}[1]{#1}
\csname url@samestyle\endcsname
\providecommand{\newblock}{\relax}
\providecommand{\bibinfo}[2]{#2}
\providecommand{\BIBentrySTDinterwordspacing}{\spaceskip=0pt\relax}
\providecommand{\BIBentryALTinterwordstretchfactor}{4}
\providecommand{\BIBentryALTinterwordspacing}{\spaceskip=\fontdimen2\font plus
\BIBentryALTinterwordstretchfactor\fontdimen3\font minus \fontdimen4\font\relax}
\providecommand{\BIBforeignlanguage}[2]{{%
\expandafter\ifx\csname l@#1\endcsname\relax
\typeout{** WARNING: IEEEtran.bst: No hyphenation pattern has been}%
\typeout{** loaded for the language `#1'. Using the pattern for}%
\typeout{** the default language instead.}%
\else
\language=\csname l@#1\endcsname
\fi
#2}}
\providecommand{\BIBdecl}{\relax}
\BIBdecl

\bibitem{TCM}
J.-L. Tang, B.-Y. Liu, and K.-W. Ma, ``Traditional chinese medicine,'' \emph{The Lancet}, vol. 372, no. 9654, pp. 1938--1940, 2008.

\bibitem{Biancang}
S.~Wei, X.~Peng, Y.-f. Wang, J.~Si, W.~Zhang, W.~Lu, X.~Wu, and Y.~Wang, ``Biancang: A traditional chinese medicine large language model,'' \emph{arXiv preprint arXiv:2411.11027}, 2024.

\bibitem{lingdan}
R.~Hua, X.~Dong, Y.~Wei, Z.~Shu, P.~Yang, Y.~Hu, S.~Zhou, H.~Sun, K.~Yan, X.~Yan \emph{et~al.}, ``Lingdan: enhancing encoding of traditional chinese medicine knowledge for clinical reasoning tasks with large language models,'' \emph{Journal of the American Medical Informatics Association}, vol.~31, no.~9, pp. 2019--2029, 2024.

\bibitem{touvron2023llama}
H.~Touvron, L.~Martin, K.~Stone, P.~Albert, A.~Almahairi, Y.~Babaei, N.~Bashlykov, S.~Batra, P.~Bhargava, S.~Bhosale \emph{et~al.}, ``Llama 2: Open foundation and fine-tuned chat models,'' \emph{arXiv preprint arXiv:2307.09288}, 2023.

\bibitem{TCMAI1}
Z.~Song, G.~Chen, and C.~Y.-C. Chen, ``Ai empowering traditional chinese medicine?'' \emph{Chemical Science}, vol.~15, no.~41, pp. 16\,844--16\,886, 2024.

\bibitem{TCMAI3}
L.~Lu, T.~Lu, C.~Tian, X.~Zhang \emph{et~al.}, ``Ai: Bridging ancient wisdom and modern innovation in traditional chinese medicine,'' \emph{JMIR Medical Informatics}, vol.~12, no.~1, p. e58491, 2024.

\bibitem{medicalGraphRAG}
J.~Wu, J.~Zhu, Y.~Qi, J.~Chen, M.~Xu, F.~Menolascina, and V.~Grau, ``Medical graph rag: Towards safe medical large language model via graph retrieval-augmented generation,'' \emph{arXiv preprint arXiv:2408.04187}, 2024.

\bibitem{GraphRAG4}
D.~Edge, H.~Trinh, N.~Cheng, J.~Bradley, A.~Chao, A.~Mody, S.~Truitt, D.~Metropolitansky, R.~O. Ness, and J.~Larson, ``From local to global: A graph rag approach to query-focused summarization,'' \emph{arXiv preprint arXiv:2404.16130}, 2024.

\bibitem{TCMcollections}
B.~H. May, C.~Lu, and C.~C. Xue, ``Collections of traditional chinese medical literature as resources for systematic searches,'' \emph{The Journal of Alternative and Complementary Medicine}, vol.~18, no.~12, pp. 1101--1107, 2012.

\bibitem{liu2024deepseek}
A.~Liu, B.~Feng, B.~Xue, B.~Wang, B.~Wu, C.~Lu, C.~Zhao, C.~Deng, C.~Zhang, C.~Ruan \emph{et~al.}, ``Deepseek-v3 technical report,'' \emph{arXiv:2412.19437}, 2024.

\bibitem{team2025kimi}
K.~Team, A.~Du, B.~Gao, B.~Xing, C.~Jiang, C.~Chen, C.~Li, C.~Xiao, C.~Du, C.~Liao \emph{et~al.}, ``Kimi k1. 5: Scaling reinforcement learning with llms,'' \emph{arXiv preprint arXiv:2501.12599}, 2025.

\bibitem{electronics13071395}
Y.~Zhang and Y.~Hao, ``Traditional chinese medicine knowledge graph construction based on large language models,'' \emph{Electronics}, vol.~13, no.~7, 2024.

\bibitem{article}
Y.~Duan, Q.~Zhou, Y.~Li, C.~Qin, Z.~Wang, H.~Kan, and J.~Hu, ``Research on a traditional chinese medicine case-based question-answering system integrating large language models and knowledge graphs,'' \emph{Frontiers in Medicine}, vol.~11, 01 2025.

\bibitem{fu2025shadead}
H.~Fu, H.~{Chen}, S.~Lin, and G.~Xing, ``{SHADE-AD}: An llm-based framework for synthesizing activity data of {Alzheimer’s} patients,'' in \emph{Proceedings of the 23rd ACM Conference on Embedded Networked Sensor Systems (SenSys'25)}.\hskip 1em plus 0.5em minus 0.4em\relax ACM, 2025.

\bibitem{qibo}
\BIBentryALTinterwordspacing
H.~Zhang, X.~Wang, Z.~Meng, Z.~Chen, P.~Zhuang, Y.~Jia, D.~Xu, and W.~Guo, ``Qibo: A large language model for traditional chinese medicine,'' 2024. [Online]. Available: \url{https://arxiv.org/abs/2403.16056}
\BIBentrySTDinterwordspacing

\bibitem{zhongjing}
S.~Yang, H.~Zhao, S.~Zhu, G.~Zhou, H.~Xu, Y.~Jia, and H.~Zan, ``Zhongjing: Enhancing the chinese medical capabilities of large language model through expert feedback and real-world multi-turn dialogue,'' in \emph{Proceedings of the AAAI conference on artificial intelligence}, vol.~38, no.~17, 2024, pp. 19\,368--19\,376.

\bibitem{tcmchat}
Y.~Dai, X.~Shao, J.~Zhang, Y.~Chen, Q.~Chen, J.~Liao, F.~Chi, J.~Zhang, and X.~Fan, ``Tcmchat: A generative large language model for traditional chinese medicine,'' \emph{Pharmacological Research}, vol. 210, p. 107530, 2024.

\bibitem{graphRAG}
B.~Peng, Y.~Zhu, Y.~Liu, X.~Bo, H.~Shi, C.~Hong, Y.~Zhang, and S.~Tang, ``Graph retrieval-augmented generation: A survey,'' \emph{arXiv preprint arXiv:2408.08921}, 2024.

\bibitem{graphRAG3}
H.~Han, Y.~Wang, H.~Shomer, K.~Guo, J.~Ding, Y.~Lei, M.~Halappanavar, R.~A. Rossi, S.~Mukherjee, X.~Tang \emph{et~al.}, ``Retrieval-augmented generation with graphs ({GraphRAG}),'' \emph{arXiv:2501.00309}, 2024.

\bibitem{chatgpt4}
J.~Achiam, S.~Adler, S.~Agarwal, L.~Ahmad, I.~Akkaya, F.~L. Aleman, D.~Almeida, J.~Altenschmidt, S.~Altman, S.~Anadkat \emph{et~al.}, ``Gpt-4 technical report,'' \emph{arXiv preprint arXiv:2303.08774}, 2023.

\bibitem{wu2023comparative}
S.~Wu, M.~Koo, L.~Blum, A.~Black, L.~Kao, F.~Scalzo, and I.~Kurtz, ``A comparative study of open-source large language models, gpt-4 and claude 2: Multiple-choice test taking in nephrology,'' \emph{arXiv preprint arXiv:2308.04709}, 2023.

\bibitem{li2024emomarker}
Y.~Li, D.~S.~F. Yu, S.~Chen, G.~Xing, and H.~Chen, ``Emomarker: A privacy-preserving, multi-modal sensing system for dyadic digital biomarkers of expressed emotions for patients with dementia,'' in \emph{Proceedings of the 22nd Annual International Conference on Mobile Systems, Applications and Services}, 2024, pp. 614--615.

\bibitem{drhouse}
B.~Yang, S.~Jiang, L.~Xu, K.~Liu, H.~Li, G.~Xing, H.~Chen, X.~Jiang, and Z.~Yan, ``Drhouse: An {LLM}-empowered diagnostic reasoning system through harnessing outcomes from sensor data and expert knowledge,'' \emph{Proceedings of the ACM on Interactive, Mobile, Wearable and Ubiquitous Technologies}, vol.~8, no.~4, pp. 1--29, 2024.

\bibitem{Retcare}
Z.~Wang, Y.~Zhu, J.~Gao, X.~Zheng, Y.~Zeng, Y.~He, B.~Jiang, W.~Tang, E.~M. Harrison, C.~Pan \emph{et~al.}, ``Retcare: Towards interpretable clinical decision making through llm-driven medical knowledge retrieval,'' in \emph{Artificial Intelligence and Data Science for Healthcare: Bridging Data-Centric AI and People-Centric Healthcare}, 2024.

\bibitem{Myotrainer_Demo}
Y.~He, X.~Wang, M.~Yuan, D.~Duan, D.~S.~F. Yu, G.~Xing, and H.~Chen, ``Myotrainer: Muscle-aware motion analysis and feedback system for in-home resistance training,'' in \emph{Proceedings of the 22nd ACM Conference on Embedded Networked Sensor Systems (SenSys'24)}.\hskip 1em plus 0.5em minus 0.4em\relax ACM, 2024.

\end{thebibliography}

\end{document}